\documentclass[a4paper, aps, pra, longbibliography, twocolumn]{revtex4-1}
\usepackage{lingmacros}
\usepackage[pagewise,switch,displaymath,mathlines]{lineno}

\usepackage{amsmath,amsfonts,amssymb}
\usepackage{graphicx}
\usepackage{subfigure}
\usepackage{color}
\usepackage{xcolor}
\definecolor{red}{rgb}{1,0,0}
\usepackage[colorlinks=true, linkcolor=blue, urlcolor=magenta, citecolor=red]{hyperref}
\newcommand{\refb}[1]{(\ref{#1})}

\newcommand{\bw}{\begin{widetext}}
\newcommand{\ew}{\end{widetext}}
\newcommand{\be}{\begin{equation}}
\newcommand{\en}{\end{equation}}
\newcommand{\bee}{\begin{equation}}
\newcommand{\ene}{\end{equation}}
\newcommand{\bea}{\begin{eqnarray}}
\newcommand{\ena}{\end{eqnarray}}

\def\pslash{p\!\!\!\slash }

\def\to{\rightarrow}

\def\pslash{p\!\!\!\slash }

\def\ie{{\it i.e.}}

\begin{document}

\title{Constrains of Charge-to-Mass Ratios on Noncommutative Phase Space}

\author{Kai Ma}
\email[Electronic address: ]{makainca@yeah.net}
\affiliation{School of Physics Science, Shaanxi University of Technology, Hanzhong 723000, Shaanxi, China}

\begin{abstract}
Based on recent measurements on the charge-to-mass ratios of proton and anti-proton, we study constraints on the parameters of noncommutative phase space. We find that while the limit on the parameter of coordinate noncommutativity is weak, it is very strong on the parameter of momentum noncommutativity, $\sqrt{\xi} \lesssim {\rm 1\mu eV}$. Therefore, the charge-to-mass ratio experiment has a strong sensitivity on the momentum noncommutativity, and enhancement of future experimental achievement can further pin down the momentum noncommutativity.
\end{abstract}


\maketitle 


\section{Introduction}\label{sec:intro}
The noncommutative filed theories are established on a noncommutative space which is characterized by a deformed algebra between coordinate operators, 
\bee\label{eq:ncdefine}
[x_{\mu}, x_{\nu}] = i \theta_{\mu\nu}\,,
\ene
and parameterized by the totally anti-symmetric constant tensor $\theta_{\mu\nu}$ which has dimension of length-squared. Such a model was originally proposed to address the infinity problem in quantum field theory~\cite{Snyder:1946qz,Yang:1947ud}, and was shown later that similar property can also appear both in string theory embedded in a background magnetic field~\cite{Seiberg:1999vs}, and quantum gravity~\cite{Freidel:2005me}.
It has been shown that the rotational symmetry can be broken~\cite{Douglas:2001ba, Szabo:2001kg}, and consquentlly the energy levels of hydrogen atom~\cite{Chaichian:2000si} and Rydberg atoms~\cite{Zhang:2004yu}, and topological phase effects~\cite{Ma:2014tua,Ma:2016rhk,Anacleto:2016ukc} as well as the quantum speed of relativistic charged particles~\cite{Wang:2017azq,Wang:2017arq,Deriglazov:2015zta,Deriglazov:2015wde} and fluid~\cite{Das:2016hmc} can receive interesting corrections. The algebra \eqref{eq:ncdefine} can be accomplished by a replacement $x_{\mu} \to x_{\mu} + p_{\mu}/(2\hbar)$. However it has been pointed out that this simple shift method can not lead to gauge invariant results~\cite{Chaichian:2008gf,Bertolami:2015yga}, and the nontrivial gauge invariant physical effects exist only for the noncommutative algebra in the momentum space described as follows~\cite{Langmann:2002cc, Langmann:2003if, Li:2006cv, Li:2006pi, Muthukumar:2006ab, Liang:2014ija,Ma:2017fnt,Ma:2016vac},
\bee\label{eq:npdefine}
[p_{\mu}, p_{\nu}] = i \xi_{\mu\nu}\,,
\ene
where $\xi_{\mu\nu}$ is also a totally anti-symmetric constant tensor, and parameterizing the momentum noncommutativity. In consideration of that the momentum operators are defined as the derivatives of the action with respect to the noncommutative coordinates, the algebra \refb{eq:npdefine} can appear naturally as a result of the algebra \refb{eq:ncdefine}.  A more general investigation on the whole Poincare group was conducted recently in Ref.\cite{Meljanac:2016jwk}, and relativistic corrections to the algebra of position variables and spin-orbital interaction were studied in Ref.~\cite{Deriglazov:2016mhk}. Therefore, in case of that the gauge problem can be cured by using the Seiberg-Witten (SW) map~\cite{Seiberg:1999vs, Ma:2014tua, Ma:2016rhk}, it is interesting to study the physical effects when both the nontrivial algebras \refb{eq:ncdefine} and \eqref{eq:npdefine} exist.

On the other hand, noncommutative field theories can invalidate the $CPT$ theorem~\cite{Chaichian:2011fc, Chaichian:2012ga,SheikhJabbari:2000vi, Chaichian:2002vw}, which is one of the most profound symmetry implied in any local and Lorentz invariant field theory~\cite{Lueders:1992dq}.
Generally, the $CPT$ invariance can be broken in extended field theories with either broken~\cite{Ellis:1992pm, Ellis:1995xd, Kostelecky:2007qf, Greenberg:2002uu, DelCima:2012gb} or conserved Lorentz symmetry~\cite{Duetsch:2012sd, Chaichian:2011fc, Chaichian:2012ga}.
The noncommutative extensions are of particular interesting since the $CPT$ symmetry can be broken in both ways~\cite{Chaichian:2011fc, Chaichian:2012ga,SheikhJabbari:2000vi}, and has been extensively studied~\cite{Chaichian:2011fc, Chaichian:2012ga,SheikhJabbari:2000vi, Chaichian:2002vw}.

In this paper, we study the constraint on the noncommutative parameters $\theta$ and $\xi$ by using recent experimental results on the charge-to-mass ratio of proton and anti-proton~\cite{Ulmer:2015jra}. The $CPT$ invariance implies proton and anti-proton have completely the same charge-to-mass ratios, apart from a sign. The measurement in Ref.~\cite{Ulmer:2015jra} gives a strong limit on possible derivation,
\bee\label{eq:exp}
\frac{ (Q/m)_{\bar{p}} }{ (Q/m)_{ p } } - 1 = 1(69) \times 10^{-12}\,.
\ene
Due to that the noncommutativities of phase space are purely geometrical properties, its physical effects are independent of composition of the particle. 
The measurements on proton and anti-proton in Ref.~\cite{Ulmer:2015jra} is expected to give strong constraints on the noncommutative parameters.

The contents of this paper is organized as follows: in Sec.~\ref{sec:nc:cf} we study the noncommutative corrections on cyclotron frequency of a charged particle in an external magnetic field; in Sec.~\ref{sec:nc:ctmr} we study the constraints of the results in Ref.~\cite{Ulmer:2015jra} on noncommutative parameters; in Sec.~\ref{sec:lv:ctmr} we study the constraints of the results in Ref.~\cite{Ulmer:2015jra} on a related model with Lorentz violation; summary is given in the final section~\ref{sec:conc}.

\section{Noncommutative Corrections on Cyclotron Frequency}\label{sec:nc:cf}
In general there are two distinct proper ``fundamental" representations for matter fields under the noncommutative $U(1)$ group~\cite{SheikhJabbari:2000vi}. The first one which is called the positive representation has a gauge transformation $\psi_{+}'=U(x)\star\psi_{+}$, while the second one which is called the negative representation is $\psi_{-}'=\psi_{-}\star U^{-1}(x)$, where the $\star$-product is a realization of the algebra \eqref{eq:ncdefine}. And for every one, there is a corresponding ``covariant" derivative, 
\bea
D^{+}_{\mu}\psi &=& \partial_{\mu}\psi+ie\psi\star A_{\mu}\,, \label{posi-repre}
  \\[2mm]
D^{-}_{\mu}\psi &=& \partial_{\mu}\psi-ie A_{\mu}\star\psi\,. \label{nega-repre}
\ena
With each of the covariant derivatives defined above, the Lagrangian
\begin{equation}\label{lagrangian}
     \mathcal{L}^{\pm}
    =\overline{\psi_{\pm}}\star(i\gamma^{\mu}D^{\pm}_{\mu}-m) \star \psi_{\pm}
\end{equation}
is invariant under the noncommutative $U(1)$ transformations. These two types of fermions are related by a charge conjugation transformation~\cite{SheikhJabbari:2000vi}. With the assumption of that the noncommutative parameter reverse its sign, \ie, $\theta \to -\theta$, under $C$ transformation, then the noncommutative  quantum electrodynamics (NCQED) preserves $C$ symmetry. Even through the above $\theta$ transformation property has an intuitive explanation~\cite{SheikhJabbari:2000vi, SheikhJabbari:1999iw, SheikhJabbari:1999vm}, it is more interesting to investigate the phenomenology of the $C$ violating NCQED in consideration of that in this case even neutral particles can couple to photons~\cite{Luo:2014iha, Wang:2017ecq}. 

No matter which representation is chosen, the SW map~\cite{Seiberg:1999vs} can be employed to keep the original gauge symmetry~\cite{Luo:2014iha, Wang:2017ecq}. It has also been shown that the coordinate and momentum noncommutativies can appear simultaneouslly in a consistent way in which the ordinary electromagnetic gauge symmetry can be preserved~\cite{Ma:2017fnt,Ma:2016vac}, and the Lagrangian density for charged particle interacting with external electromagnetic fields has been obtained as follows, 
\bee\label{eq:lagrangian}
\mathcal{L} =
\bar{\psi}(x)( \pslash - Q ~\slash{\!\!\!\!A}_{NC} - m ) \psi(x) \,,
\ene
where $Q$ is the charge of matter particle in unite of $|e|$, and the effective potential 
$A_{NC;\mu} = A_{\mu} + A_{\xi;\mu}$, \ie, a sum of the original one $A_{\mu}$ and an effective term $A_{\xi;\mu}$ emerging due to the noncommutativity of momenta operators and having following expression,
\bee\label{eq:Ashift}
A_{\xi;\mu} = \frac{\xi }{2\hbar Q} \big(0, y,\, -x,\, 0  \big)\,.
\ene
It should be stressed that the above expression of $A_{\xi;\mu}$ is obtained by defining  
the $\hat{z}$-axis as the direction of the vector $\vec{\xi}$ whose components are required to related with the noncommutative parameter $\theta_{ij}$ by the relation $\xi_{i}=\epsilon_{ijk}\xi^{ij}/2$ such that non-zero components are only
$\xi^{12}=-\xi^{21}=\xi$. In this configuration, the effect of momentum-momentum 
noncommutativity is an addition of a constant magnetic background field
$
\vec{B}_{\xi} 
= \vec{\nabla} \times \vec{A}_{\xi} 
= \big(0, 0, B_{\xi}  \big)
$
, with $B_{\xi} = \xi /(\hbar Q)$  over the whole space. The non-relativistic approximation can be obtained by using the well-known Foldy-Wouthuysen unitary transformation (FWUF)~\cite{Foldy:1949wa}, and neglecting the spin degree of freedom the non-relativistic Hamiltonian is given as~\cite{Ma:2017fnt,Ma:2016vac},
\begin{equation}\label{eq:nonRelNCH}
H_{NC} = \frac{1}{2m_{\theta}}(\vec{p} - Q\vec{A}_{NC})^2\,,
\end{equation}
where the noncommutative effective mass 
\bee\label{eq:massren}
m_{\theta} = m\alpha_{\theta}^{-1}\,,\;\;\; 
\alpha_{\theta} = 1- Q \frac{ \vec{\theta}\cdot\vec{B}_{NC} }{2\phi_{0}}\,,
\ene
and $\phi_{0}=h/e$ is the fundamental magnetic flux. Because of that $|\theta||\xi| \ll \hbar^2$, and furthermore usually the external magnetic field $|\vec{B}|$ is much stronger then the noncommutative background $|\vec{B}_{\xi}|$, the scale factor $\alpha_{\theta}$ in  in \eqref{eq:massren} can be approximated as 
\bee
\alpha_{\theta} \approx 1- Q \frac{ \vec{\theta}\cdot\vec{B}}{2\phi_{0}} \,.
\ene 
We will use this approximation in the rest of this paper to estimate the constraint on noncommutative parameters.

The charge-to-mass ratios reported in Ref.~\cite{Ulmer:2015jra} was obtained by measuring the cyclotron frequency of charged particles in a constant external magnetic field $B_{0}=1.946{\rm T}$. Therefore we need to know the dynamical properties of charged particle in a constant external magnetic field on noncommutative phase space. We require that the external magnetic field is also along the $z$ direction, \ie, $\vec{B}=B\vec{e}_{z}$. This is not true in general, but is a good approximation since that the noncommutative correction is maximum in this case. On the other hand the measurement on the sidereal variations in Ref.~\cite{Ulmer:2015jra}, which gives an upper bound of $720$ parts per trillion that is a little weaker then \eqref{eq:exp}, also justifies our approximation. Furthermore, we chose the symmetric gauge to solve the static Schr\"odinger equation, and the ordinary gauge potential can be expressed in this gauge as, $\vec{A} = \frac{1}{2}B(-y, x, 0)$. In consideration of that the $z$ component can be factorized completely and plane wave solutions are sought, we will neglect it in the rest of this paper, and explicit expression of the transverse part of the Hamiltonian  \eqref{eq:nonRelNCH} can be obtained from upon expansion,
\bee
H_{\|} 
= \frac{p_{x}^{2}}{2m_{\theta}} + \frac{1}{2}m_{\theta} \omega^{2}x^{2}
+ \frac{p_{y}^{2}}{2m_{\theta}} + \frac{1}{2}m_{\theta} \omega^{2}y^{2}\,,
\ene
where the Larmor frequency $\omega = QB_{NC} /(2m_{\theta})$. One can see that this Hamiltonian is mimic to the ordinary Landau problem, apart from a correction on the Larmor frequency. The noncommutative extension of the Landau system have been studied extensively~\cite{Horvathy:2002wc, Dulat:2008eu, Giri:2008qu, Alvarez:2009nz, Maceda:2009zz, Fiore:2011kh, Luo:2013kka, Gangopadhyay:2014afa, DIAOXin-Feng:2015fra}. However, so far the charge-to-mass ratio related physics have not been studied. It is well-known that the corresponding eigenvalue problem can be solved exactly in terms of polar coordinates, and the energy eigenvalues are given as,
\bee
E = ( n  + \frac{1}{2}) \hbar\omega_{C}\,,\;\;\; \omega_{C}=2\omega\,,
\ene
where $\omega_{C}$ is the cyclotron frequency.

\section{Constraints of Charge-to-Mass Ration}\label{sec:nc:ctmr}
The charge-to-mass ratio can be extracted from the measured cyclotron frequency and external magnetic field as follows,
\bee
\bigg[ \frac{Q}{m}  \bigg]_{\rm exp} 
=  \frac{ \omega_{C} }{ B }
= \frac{Q}{m} \cdot  \alpha_{\theta} \cdot \frac{ B_{NC} }{ B }\,.
\ene
In case of that noncommutative parameters are small, the noncommutative corrections can be approximated as
\bee
\bigg[ \frac{Q}{m}  \bigg]_{\rm exp}  
\approx 
\frac{Q}{m} \bigg( 1- \frac{ Q \theta B }{2\phi_{0}}  + \frac{ \xi }{ \hbar Q B }  \bigg)\,.
\ene
Under this approximation, the antiproton-to-proton mass ratio is given as
\bee
\chi_{NC} 
= \frac{ (Q/m_{\theta})_{\bar{p}} }{ (Q/m_{\theta})_{ p } }
\approx \chi + \frac{ \theta B }{\phi_{0}}  - \frac{  2 \xi }{ e\hbar B } \,,
\ene
where $\chi=1$ due to the $CPT$ conservation of the ordinary field theory. Therefore, by require the noncommutative corrections lie in the $1\sigma$ region of the experimental error, the result \eqref{eq:exp} given in Ref.~\cite{Ulmer:2015jra} puts a constraint
\bee
\bigg| \frac{ \theta B_{0} }{\phi_{0}}  - \frac{  2 \xi }{ e\hbar B_{0} } \bigg| \le 0.69\times 10^{-12}
\ene
on the noncommutative parameter $\theta$ and $\xi$. In case of that either $\theta$ or $\xi$ vanish, one has following upper limits,
\bea
\theta &\le& 1.46 \times 10^{-27} {\rm m^{2}} \,,
\\[2mm]
\xi &\le& 1.14 \times 10^{-65} {\rm kg^{2}\cdot m^{2} \cdot s^{-2} } \,.
\ena
While the limit on the noncommutative parameter $\theta$ is not strong, $1/\sqrt{\theta}\gtrsim 0.5 {\rm MeV}$ in unite of energy, the constraint on the noncommutative parameter $\xi$ is very strong, $\sqrt{\xi} \lesssim {\rm 1\mu eV}$. Therefore, the charge-to-mass ratio experiment performed in in Ref.~\cite{Ulmer:2015jra} has a strong sensitivity on the momentum noncommutativity.

\section{Constraint on Lorentz Violation Parameter}\label{sec:lv:ctmr}
It has been pointed out that, the noncommutative extension of quantum field theory can be effectively described by a quantum field theory with Lorentz violation~\cite{Chaichian:2004za,Carroll:2001ws}. Therefore, it is expected to give  give strong constraints on the Lorentz violation parameters. In this section, we study constraint of charge-to-mass ratio on the Lorentz violation. 

In general there can be a lot of parameters in a quantum field theory with Lorentz violation~\cite{Husain:2015tna, Balachandran:2014hra}. Here we consider only following Lagrangian, 
\begin{equation}\label{lagrangian:lv}
\mathcal{L}
=\overline{\psi}(i\gamma^{\mu}D_{\mu} + iQc^{\mu\nu}\gamma_{\mu}D_{\nu}-m)\psi\,,
\end{equation}
where $c^{\mu\nu}$ is a constant tensor, and parameterizing the strength of Lorentz violation. The CPT invariance is explicitly violated by the charge dependence of the anomalous interaction term. The measurement in Ref.~\cite{Ulmer:2015jra} is expected to give a strong limit on the parameter $c^{\mu\nu}$.
The non-relativistic approximation can be obtained by using the well-known Foldy-Wouthuysen unitary transformation (FWUF)~\cite{Foldy:1949wa}, and neglecting the spin degree of freedom the non-relativistic Hamiltonian is given as~\cite{Goncalves:2014jwa},
\begin{equation}\label{eq:nonRelNCH}
H = \frac{1}{2\tilde{m}}(\vec{p} - Q\vec{A})^2\,,
\end{equation}
where the effective mass 
\bee\label{eq:massren}
\tilde{m} = m(1+Qc_{00})\,.
\ene
Neglecting the trivial dynamics along $z$ direction, the explicate expression of the transverse part of the Hamiltonian \eqref{eq:nonRelNCH} is,
\bee
H_{\|} 
= \frac{p_{x}^{2}}{2\tilde{m}} + \frac{1}{2}\tilde{m} \omega^{2}x^{2}
+ \frac{p_{y}^{2}}{2\tilde{m}} + \frac{1}{2}\tilde{m} \omega^{2}y^{2}\,,
\ene
where the Larmor frequency $\omega = QB /(2\tilde{m})$. One can see that this Hamiltonian is mimic to the ordinary Landau problem. The eigenvalue problem can then be solved exactly in terms of polar coordinates, and the energy eigenvalue is given as follows,
\bee
E = ( n  + \frac{1}{2}) \hbar\omega_{C}\,,\;\;\; \omega_{C}=2\omega\,,
\ene
where $\omega_{C}$ is the cyclotron frequency.

The charge-to-mass ratio can be extracted from the measured cyclotron frequency and external magnetic field as follows,
\bee
\bigg[ \frac{Q}{m}  \bigg]_{\rm exp} 
=  \frac{ \omega_{C} }{ B } 
= \frac{Q}{m(1 + Qc_{00})} \,.
\ene
In case of that parameters $c_{00}$ are small, the corrections can be approximated as,
\bee
\bigg[ \frac{Q}{m}  \bigg]_{\rm exp}  
\approx 
\frac{Q}{m} (1 - Qc_{00})\,.
\ene
In this approximation, the antiproton-to-proton mass ratio is given as,
\bee
\tilde{\chi} 
= \frac{ (Q/m_{\theta})_{\bar{p}} }{ (Q/m_{\theta})_{ p } }
\approx \chi + 2c_{00} \,,
\ene
where $\chi=1$ due to the CPT conservation of the ordinary field theory. Therefore, by require the corrections lie in the $1\sigma$ region of the experimental error, the result \eqref{eq:exp} given in Ref.~\cite{Ulmer:2015jra} give a following constraint,
\bee
\bigg| c_{00} \bigg| \le 0.345\times 10^{-12}\,.
\ene

\section{Summary}\label{sec:conc}

In summary, we study the quantum properties of a charged particle in a constant external magnetic field, and by using the recent measurement on the charge-to-mass ratios of proton and anti-proton~\cite{Ulmer:2015jra}, we have shown that while the charge-to-mass ratio experiment is not sensitive to the parameter of coordinate noncommutativity, it can give strong constraint on the parameter of momentum noncommutativity. The current bound is $\sqrt{\xi} \lesssim {\rm 1\mu eV}$ (in unite of energy). It is expected that future enhancement of experimental precision can further pin down the momentum noncommutativity. We also studied related model with Lorentz violation.

\section*{Acknowledgements}
K.M. is supported by the China Scholarship Council, and the National Natural Science Foundation of China under Grant No. 11647018 and 11705113.

\bibliography{aString}
\end{document}